\documentclass[10pt,journal,twocolumn]{IEEEtran}
\IEEEoverridecommandlockouts
\usepackage{cite}
\usepackage{amsmath,amssymb,amsfonts,booktabs}
\usepackage{algorithmic}
\usepackage{graphicx}
\usepackage{textcomp}
\usepackage{xcolor}
\usepackage[table]{xcolor}
\usepackage{enumerate}
\usepackage{epstopdf}
\usepackage{subcaption}
\usepackage{amsmath,amsfonts}
\usepackage{amsthm}
\usepackage{amssymb}
\usepackage{easyReview}
\usepackage{algorithmic}
\usepackage{algorithm}
\usepackage{array}
\usepackage[caption=false,font=normalsize,labelfont=sf,textfont=sf]{subfig}
\usepackage{easyReview}
\usepackage{textcomp}
\usepackage{stfloats}
\usepackage{url}
\usepackage{bm}
\usepackage{makecell}
\usepackage{verbatim}
\usepackage{graphicx}
\usepackage{caption}
\usepackage{subfig}
\usepackage{subcaption}
\usepackage{cite}
\usepackage{booktabs}
\usepackage{amsthm}

\usepackage{adjustbox}

\usepackage{array}
\definecolor{mygray}{RGB}{240,240,240}

\allowdisplaybreaks[4]

\def\BibTeX{{\rm B\kern-.05em{\sc i\kern-.025em b}\kern-.08em
		T\kern-.1667em\lower.7ex\hbox{E}\kern-.125emX}}
\begin{document}

\title{Fluid Antennas Enabled UAV Communication Systems: Fundamentals, Challenges,\\and Future Directions}

\author{Wenchao Liu,  
        Weijie Yuan, 
        Jun Wu,
        Jinke Ren,    
        Xuhui Zhang,
        and Changsheng You

\thanks{
Wenchao Liu, Weijie Yuan, and Jun Wu are with the School of Automation and Intelligent Manufacturing, Southern University of Science and Technology, Shenzhen 518055, China. (e-mail: wc.liu@foxmail.com; yuanwj@sustech.edu.cn; wuj2021@mail.sustech.edu.cn).
}

\thanks{
Jinke Ren and Xuhui Zhang are with the School of
Science and Engineering, Shenzhen Future Network of Intelligence Institute, and Guangdong Provincial Key Laboratory of Future Networks of
Intelligence, The Chinese University of Hong Kong (Shenzhen), Longgang, Shenzhen, Guangdong 518172, China. (e-mail: jinkeren@cuhk.edu.cn; xu.hui.zhang@foxmail.com).
}

\thanks{Changsheng You is with the Department of Electronic and Electrical Engineering, Southern University of Science and Technology, Shenzhen 518055, China. (e-mail: youcs@sustech.edu.cn).
}

\thanks{Corresponding author: Weijie Yuan}

}

\maketitle

\begin{abstract} 
Low-altitude wireless networks (LAWNs) require uncrewed aerial vehicle (UAV) communication systems to adapt to large-scale propagation changes and fast local channel fluctuations. Fluid antennas (FAs) provide an additional degree of freedom (DoF) by reconfiguring antenna positions within a compact aperture. This article investigates FA-enabled UAV communication systems from the perspectives of fundamentals, design challenges, and future research directions. We first introduce a dual-mobility perspective, in which UAV trajectory planning provides macro-mobility for large-scale air-to-ground geometry adaptation, and FA position reconfiguration provides micro-mobility for fine-grained local channel adaptation. We then identify key design challenges arising from the high computational complexity and information uncertainty in the joint optimization of UAV trajectory, FA positions, and beamforming under mobility, energy, hardware, and security constraints. To address these challenges, we present a hybrid offline-online optimization framework, where prior information such as environment maps, user mobility patterns, and statistical channel state information (CSI) supports long-term planning, while real-time information such as instantaneous CSI and UAV state enables fast online correction. A case study on FA-enabled UAV-assisted vehicular communications demonstrates the effectiveness of the proposed framework. Finally, we discuss future directions of FA-enabled UAV communication in integrated sensing and communication (ISAC), secure communication, mobile edge computing (MEC), and aerial relaying. 
\end{abstract}

\section{Introduction}
Uncrewed aerial vehicles (UAVs) have emerged as promising platforms for low-altitude wireless networks (LAWNs), owing to their flexible deployment, three-dimensional (3D) mobility, and ability to establish air-to-ground (A2G) links~\cite{11503149}. 
Compared with terrestrial base stations (BSs), UAVs can dynamically adjust their locations and trajectories to improve propagation conditions, increase the line-of-sight (LoS) probability, reduce path loss, and avoid severe blockages caused by buildings or terrain~\cite{8698468}. 
These capabilities make UAV communication systems particularly attractive for temporary coverage, emergency communications, aerial relaying, vehicular communications, and Internet of Things (IoT) data collection.

However, UAV mobility alone is insufficient to adapt to the highly dynamic and heterogeneous low-altitude communication environments. UAV trajectory planning operates on a relatively slow timescale and is constrained by onboard energy, flight dynamics, no-fly zones, collision avoidance, and mission endurance. Consequently, UAV maneuvering cannot respond quickly to small-scale fading, sudden blockages, user mobility, attitude jitter, or rapidly varying multi-user interference~\cite{11443232}. Additional physical-layer degrees of freedom (DoFs) are therefore needed to provide fast and fine-grained channel adaptation while improving link reliability and spectral efficiency.

\begin{figure}[t]
\centering
\fbox{\includegraphics[width=0.95\linewidth]{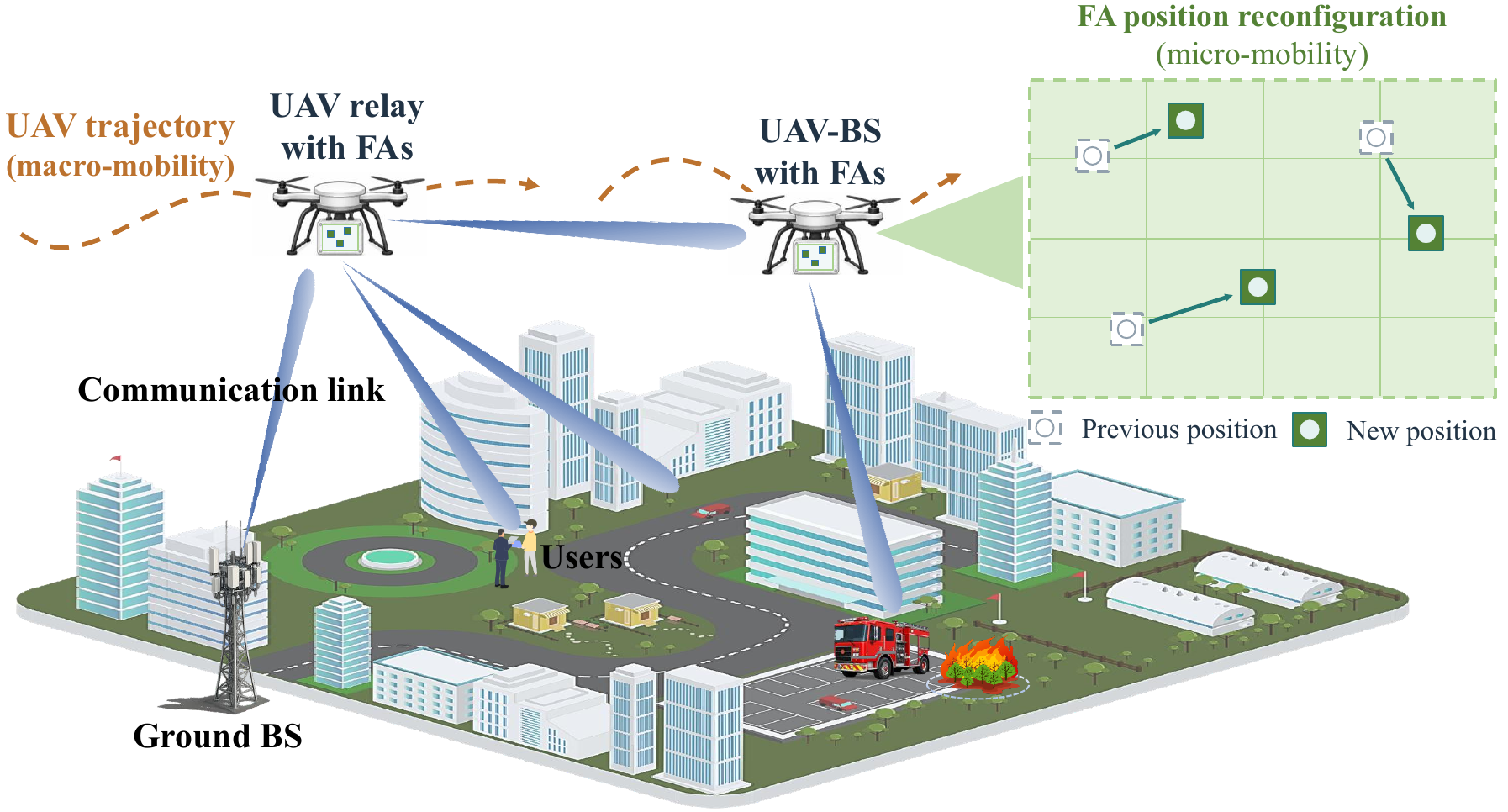}}
\caption{A representative FA-enabled UAV communication architecture in LAWNs, where UAVs serve as BSs and relays. In this architecture, UAV-mounted FAs enable local antenna-position reconfiguration.}
\label{fig:1}
\end{figure}

Fluid antennas (FAs), also known as movable antennas (MAs), provide a promising solution to this issue~\cite{10286328, 10909643}. Unlike conventional fixed-position antennas (FPAs), FAs can reconfigure the positions of radiating elements within a bounded aperture, thereby changing the effective wireless channel~\cite{10654366}. 
As illustrated in Fig.~\ref{fig:1}, an FA-enabled UAV communication system exploits two forms of mobility in a LAWN: the macro-mobility of the UAV trajectory and the micro-mobility of FA position reconfiguration. 
Their complementary integration is defined as dual mobility. UAV macro-mobility adjusts the large-scale A2G propagation geometry, including the link distance, elevation angle, and blockage condition, whereas FA micro-mobility selects favorable local channel responses through antenna-position reconfiguration. Operating over different spatial and temporal scales, these two forms of mobility are complementary and motivate the joint design of UAV trajectory, FA positions, and beamforming.
\footnote{This article focuses on UAV-mounted FAs, where trajectory mobility is combined with local antenna-position reconfiguration. FAs may also be deployed at BSs, vehicles, or other ground terminals, but such architectures involve different hardware constraints and optimization variables.}

Introducing FA positions as optimization variables transforms the conventional joint design of UAV trajectory and beamforming into a multi-variable optimization problem involving UAV trajectory, FA positions, and beamforming. Their strong coupling, together with UAV mobility limits, FA reconfiguration constraints, power and energy budgets, and security requirements, makes the joint design computationally challenging.
Moreover, slowly varying prior information, such as environment maps and user mobility patterns, and rapidly varying real-time information, such as instantaneous channel state information (CSI) and UAV state, are available over different timescales, introducing uncertainty and requiring efficient offline-online coordination.

Motivated by these observations, this article presents a system-level perspective on FA-enabled UAV communication with dual mobility as the central organizing principle. The main contributions are threefold.
First, we explain the advantages of introducing FAs into UAV communications in terms of onboard resource constraints, response lag compensation, attitude perturbation mitigation, multi-user interference management, and mission-oriented scenario adaptation.
Second, we identify the key design challenges arising from the coupling among UAV trajectory, FA positions, and beamforming, and propose a hybrid offline-online optimization framework that exploits prior information for long-term planning and real-time information for fast online adaptation. A case study on FA-enabled UAV-assisted vehicular communications demonstrates the effectiveness of this framework.
Third, we discuss promising research directions, including integrated sensing and communication (ISAC), secure communication, mobile edge computing (MEC), and aerial relaying.

\section{Fundamentals of FA-Enabled UAV Communications}
\subsection{Motivation for Introducing FAs into UAV Communications}
Building on the dual mobility introduced in Section I, we explain why FAs are particularly suitable for UAV communication systems. The key observation is that UAV trajectory planning and FA position reconfiguration operate on complementary spatial and temporal scales~\cite{11214460}. UAV trajectory planning primarily adjusts the large-scale propagation geometry, including the A2G link distance, elevation angle, and blockage condition, whereas FA position reconfiguration enables fine-grained local channel adaptation within the antenna aperture. Thus, FAs complement UAV mobility by providing a faster physical-layer adaptation capability.

This complementarity is particularly valuable in LAWNs, where UAV trajectory planning is constrained by flight dynamics, energy budgets, no-fly zones, collision avoidance, and mission duration, and therefore cannot respond rapidly to local channel variations.
In contrast, FAs can select favorable antenna positions to mitigate small-scale fading, compensate for attitude-induced channel perturbations, and reshape multi-user channel correlations~\cite{11119811}. 
These capabilities make FAs well suited to dynamic UAV communication environments.

\subsection{Advantages of FA-Enabled UAV Communication Systems}
\subsubsection{Onboard Resource Constraints}
FAs are well suited to UAV platforms with limited onboard resources. 
Conventional multi-antenna arrays achieve array, diversity, and spatial multiplexing gains by increasing the number of antenna elements and RF chains~\cite{10354003}. 
However, this approach is difficult to implement on UAVs because of payload weight and power consumption constraints.
By selecting favorable antenna positions, FAs provide additional spatial selectivity with moderate hardware complexity, offering a lightweight and reconfigurable means of exploiting spatial diversity.

\subsubsection{Response Lag Compensation}
UAV trajectory planning is governed by flight dynamics and therefore responds much more slowly than that of small-scale channel fluctuations, whereas FA position reconfiguration can operate on a shorter timescale. 
FAs can thus improve instantaneous link quality even when the UAV is hovering, following a prescribed trajectory, or restricted to a safe flight corridor~\cite{11271535}. 
This capability enables fast local channel adaptation without frequent UAV maneuvering.

\subsubsection{Attitude Perturbation Mitigation}
FAs can mitigate the impact of UAV attitude perturbations. Rotor vibration, wind gusts, maneuvering, and attitude-control errors can cause short-term variations in antenna position, polarization, phase relationships, and link geometry. With FPAs, these perturbations directly translate into equivalent-channel fluctuations, whereas FAs can adapt their positions to compensate for such variations. This position-domain adaptation can reduce received power fluctuations and coherent-combining losses, thereby improving link reliability and service continuity.

\subsubsection{Multi-User Interference Management}
FAs also provide an additional DoF for multi-user interference management~\cite{11302793}. 
Conventional beamforming optimizes transmit weights under fixed antenna positions, whereas FAs additionally allow the effective channel response to be reconfigured. Different FA positions can induce different channel correlations and interference coupling patterns, thereby changing the feasible beamforming space. Joint optimization of FA positions and beamforming can therefore improve both desired-signal enhancement and inter-user interference suppression.

\subsubsection{Mission-Oriented Scenario Adaptation}
FAs are also attractive for mission-oriented applications such as temporary coverage, emergency communications, aerial relaying, and IoT data collection~\cite{11240557}. These scenarios often involve rapid deployment, dynamic network topologies, and incomplete infrastructure. 
By enabling fine-grained local channel reconfiguration, FAs allow UAVs to adapt to changing user distributions, blockage conditions, and service requirements without frequent trajectory adjustments. 
This capability is particularly beneficial when UAVs operate under limited energy budgets, restricted flight paths, or steady-hovering requirements.

\begin{figure*}[t]
\centering
\fbox{\includegraphics[width=0.95\linewidth]{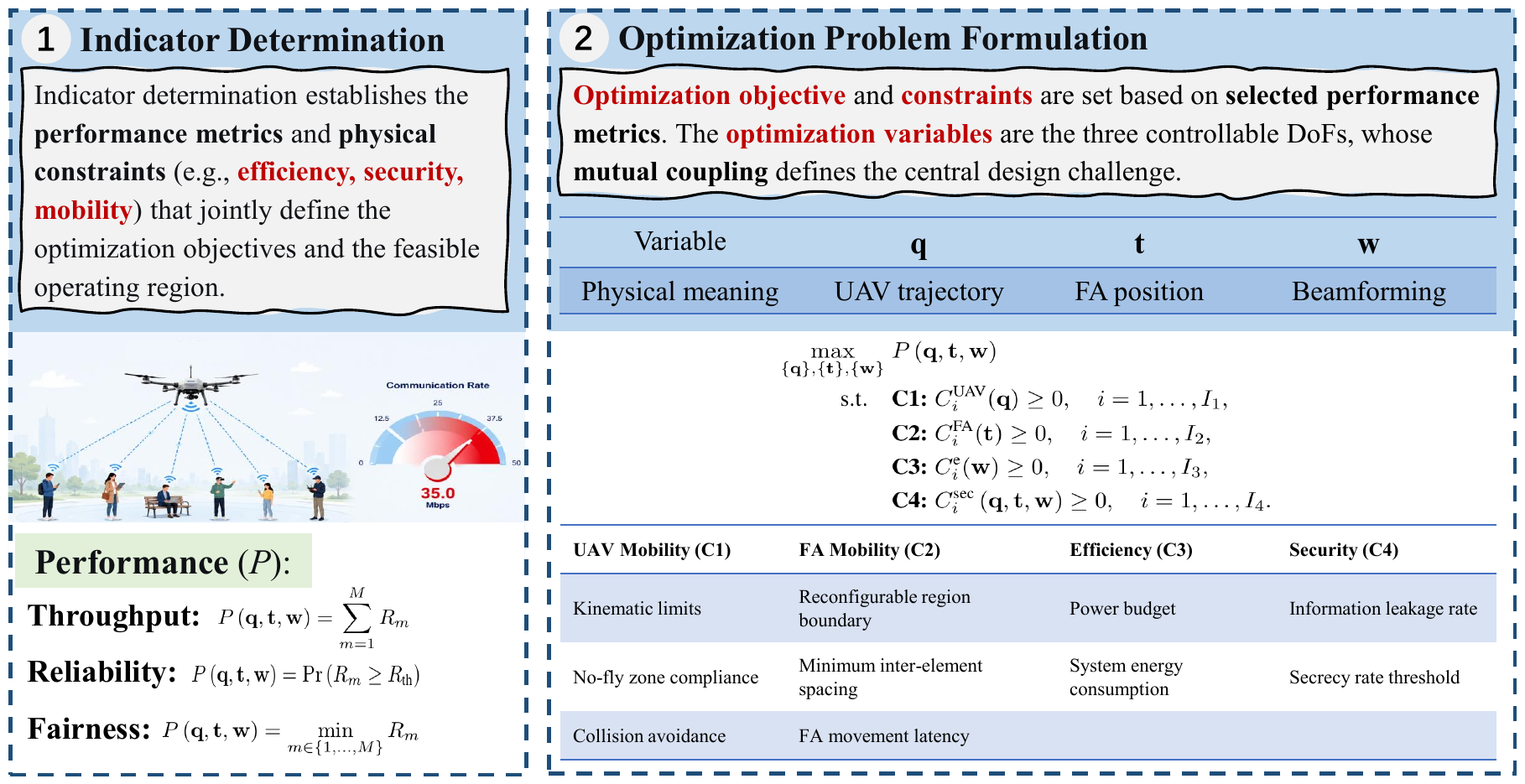}}
\caption{Joint optimization framework for FA-enabled UAV communication systems, where UAV trajectory, FA positions, and beamforming are jointly designed under performance and practical constraints.}
\label{fig:2}
\end{figure*}

\section{Design Challenges for FA-Enabled UAV Communications}
Integrating FA position reconfiguration into UAV communication systems introduces two fundamental design challenges. First, the joint optimization problem of UAV trajectory, FA positions, and beamforming is highly coupled and computationally challenging because these variables jointly determine the effective channel under mobility, hardware, energy, and security constraints. Second, slowly varying prior information and rapidly varying real-time information differ substantially in availability, accuracy, and timescale. Prior information is mainly available before the mission for long-term planning, whereas instantaneous CSI and UAV state are acquired during operation and require fast online adaptation~\cite{9034493}. 
Accordingly, Sections III-A and III-B discuss the computational complexity of multi-variable coupling and the information uncertainty and timescale mismatch, respectively.

\subsection{High Computational Complexity}
As illustrated in Fig.~\ref{fig:2}, UAV trajectory, FA positions, and beamforming are strongly coupled because they jointly determine the effective channel and communication performance. The UAV trajectory controls the large-scale propagation geometry, including the link distance, elevation angle, and LoS probability; FA position reconfiguration selects the favorable local channel responses within the reconfigurable aperture; and beamforming exploits the resulting effective channel for array gain and interference suppression.

This coupling prevents the three variables from being optimized independently. A favorable UAV trajectory with poorly selected FA positions may still yield weak effective channels and limited beamforming gains, whereas optimized FA positions cannot compensate for a trajectory that places the UAV in an unfavorable propagation region~\cite{11513706}. Moreover, the effective channel depends nonlinearly on both UAV trajectory and FA positions, while achievable-rate constraints depend on the signal-to-interference-plus-noise ratio involving all three variables. The resulting joint optimization is therefore inherently non-convex.

Practical constraints further increase the difficulty of the problem. UAV mobility is limited by speed and acceleration constraints, no-fly zones, and collision avoidance, while FA reconfiguration is restricted by the aperture boundary, minimum inter-element spacing, and repositioning latency. Power, energy, and security requirements impose additional constraints on transmit power, system energy consumption, and information leakage. Together, the variable coupling and these heterogeneous constraints lead to a highly non-convex problem that typically requires iterative optimization, making real-time implementation on UAV platforms challenging.

\subsection{Information Uncertainty and Timescale Mismatch}
The second challenge arises from the different accuracy, availability, and timescales of the prior and real-time information. 
Prior information, such as environment maps, building distributions, user mobility statistics, probabilistic LoS models, and statistical CSI, is available before the mission and evolves relatively slowly. It is therefore suitable for long-term offline planning. In contrast, instantaneous CSI, link-quality feedback, and the current UAV state are time-sensitive and more suitable for fast online adaptation~\cite{10065443}.

Although prior information provides a statistical basis for UAV trajectory, FA position, and beamforming design, it may deviate from the actual environment because of outdated environment maps, unpredictable user mobility, or inaccurate channel models. During operation, sudden blockages, user movement, attitude variations, and small-scale fading can further invalidate offline predictions. Real-time information is therefore required to correct such mismatches and maintain reliable communication.

These two types of information naturally correspond to different decision timescales. The offline stage operates at a mission-level timescale and uses prior information for long-term planning, whereas the online stage operates at a much shorter slot-level timescale and uses real-time information for local adaptation. This timescale separation motivates an offline-online design in which offline planning reduces the real-time optimization burden and online adaptation compensates for deviations from prior predictions.

\section{Hybrid Offline-Online Optimization Framework}
To address the high computational complexity and information-timescale mismatch identified in Section III, we consider a hybrid offline-online optimization framework that decomposes the multi-variable joint design according to information availability and decision timescale. The offline stage uses prior information for computationally intensive long-term planning, while the online stage exploits real-time information for low-complexity local adaptation within the structured decision space determined offline.

\begin{figure*}[t]
\centering
\fbox{\includegraphics[width=0.95\linewidth]{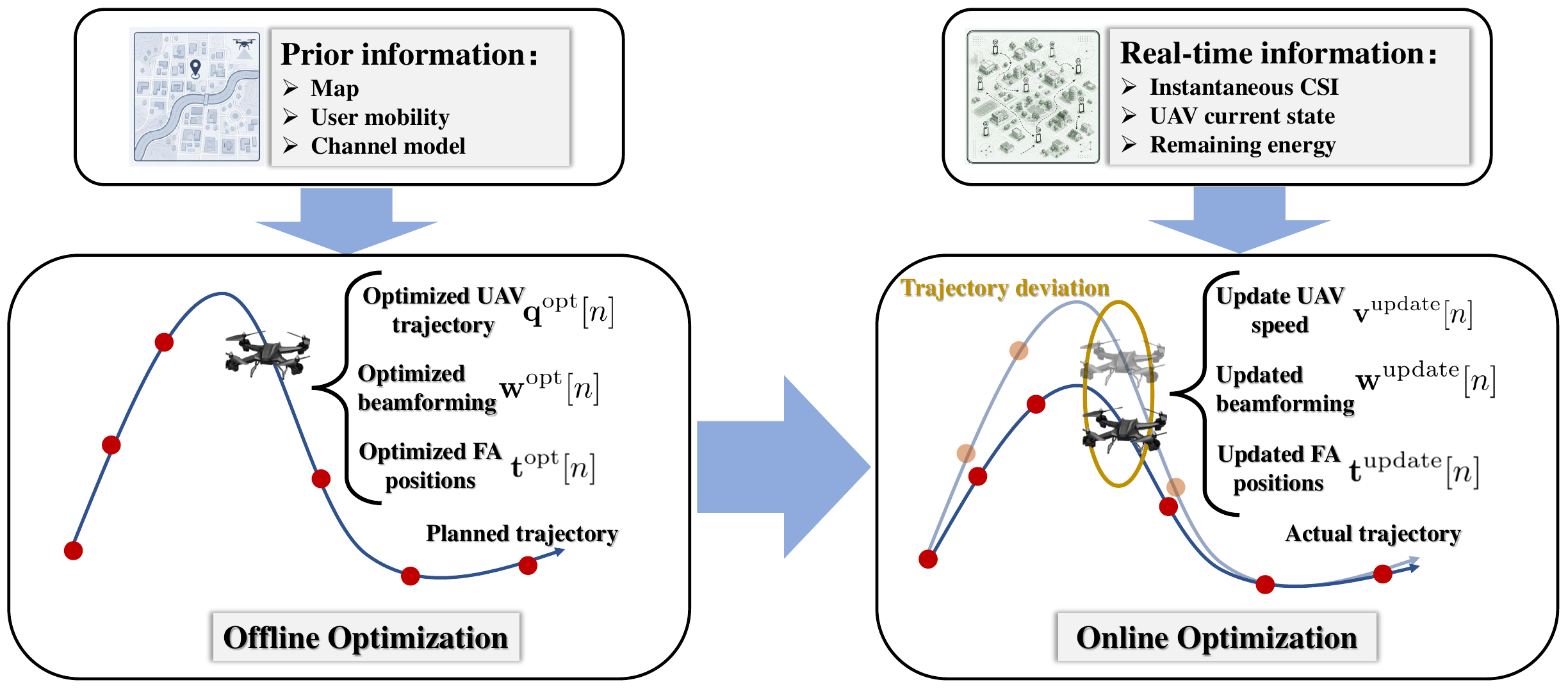}}
\caption{Information flow of the hybrid offline-online optimization framework, where offline outputs provide references and reduced candidate spaces for real-time adaptation.}
\label{fig:3}
\end{figure*}

\subsection{Framework Overview}
As illustrated in Fig.~\ref{fig:3}, the offline stage exploits prior information, including environment maps, user mobility statistics, and probabilistic channel information, to generate a reference UAV trajectory, candidate FA position sets, and reference beamforming policies or codebooks. Rather than serving as fixed decisions, these outputs provide a baseline and reduced search space for online adaptation.

During operation, the online stage uses instantaneous CSI, the current UAV state, and the remaining energy to locally refine the offline decisions. Specifically, it adjusts the UAV speed and, when necessary, permits limited trajectory deviation around the reference trajectory, selects or updates FA positions within the candidate sets, and adapts beamforming according to instantaneous CSI. Thus, the offline stage performs mission-level global planning, whereas the online stage performs slot-level local correction, balancing computational efficiency and responsiveness to channel and mobility variations.

\subsection{Offline Optimization: Long-Term Planning Based on Prior Information}
Offline optimization is performed before the mission using prior information that characterizes the long-term statistical behavior of the low-altitude environment. Environment maps and building distributions help identify LoS-dominant and blockage-prone regions, enabling the UAV to plan a reference trajectory that avoids persistently unfavorable propagation conditions. User mobility statistics indicate where service demand is likely to occur and can be used to identify candidate service segments or low-speed flight regions. Statistical channel models further characterize the expected link quality under different UAV and FA positions, allowing the candidate FA position sets and reference beamforming strategies to be designed in advance.

Accordingly, the offline stage generates three main outputs: (i) a reference UAV trajectory defining the nominal long-term flight path, (ii) candidate FA position sets that reduce the online antenna-position search space, and (iii) reference beamforming policies or codebooks that provide initial transmission strategies. These outputs form a structured decision space for online adaptation and avoid full-dimensional joint optimization during real-time operation. Since computation is completed before the mission, relatively sophisticated methods, such as alternating optimization, sequential convex approximation, and continuous relaxation, can be employed to handle the coupling among UAV trajectory, FA positions, and beamforming.

\subsection{Online Optimization: Fast Correction Based on Real-Time Information}
Online optimization operates during UAV flight using instantaneous CSI, UAV state, and remaining energy. The system first evaluates the current A2G link quality relative to the offline prediction. Under favorable propagation conditions, the UAV may retain or slightly refine the offline FA positions and beamforming strategy and reduce its speed to exploit transmission opportunities. When sudden blockage, deep channel fading, or significant link degradation occurs, the system updates the FA positions and beamforming and may increase the UAV speed to leave the unfavorable region.

Because online decisions are constrained by flight dynamics, energy, mission deadlines, and FA reconfiguration overhead, global trajectory replanning or exhaustive antenna-position search is generally impractical. The UAV therefore primarily adjusts its flight speed and, when necessary, makes limited deviations from the reference trajectory. FA positions are selected or updated within the candidate sets generated offline, while beamforming is adapted according to instantaneous CSI and the selected FA positions. The online design should also account for the control-signaling overhead and repositioning latency associated with FA reconfiguration to avoid excessive switching. By restricting real-time optimization to this local decision space, the framework substantially reduces the computational complexity while retaining the ability to respond to channel and mobility variations.

\section{Case Study}
\begin{figure}[t]
\centering
\fbox{\includegraphics[width=0.95\linewidth]{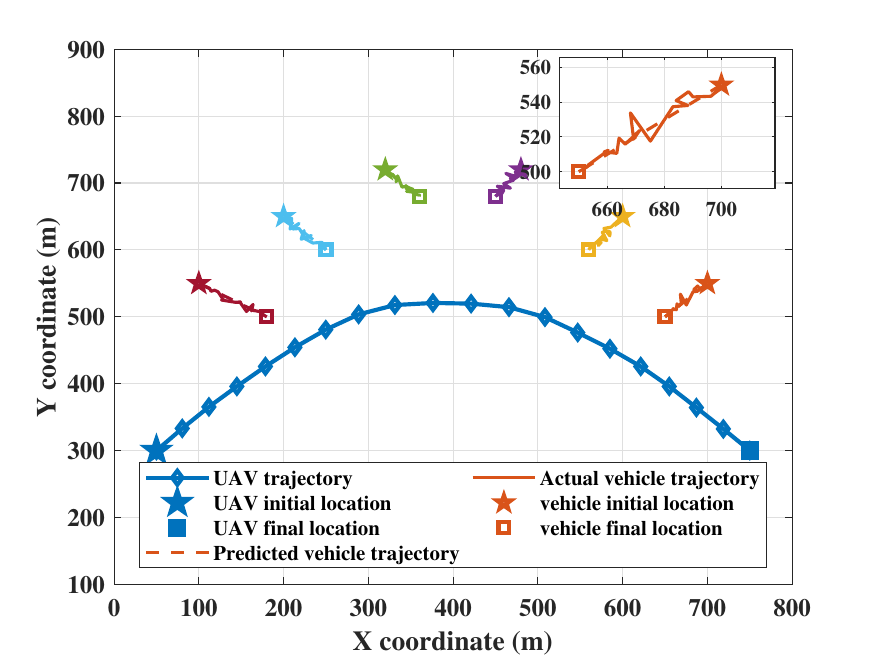}}
\caption{The UAV-assisted vehicular communication scenario, where predicted vehicle trajectories are used for offline planning and actual trajectories are observed during online adaptation.}
\label{fig:4}
\end{figure}

\begin{figure}[t]
\centering
\fbox{\includegraphics[width=0.95\linewidth]{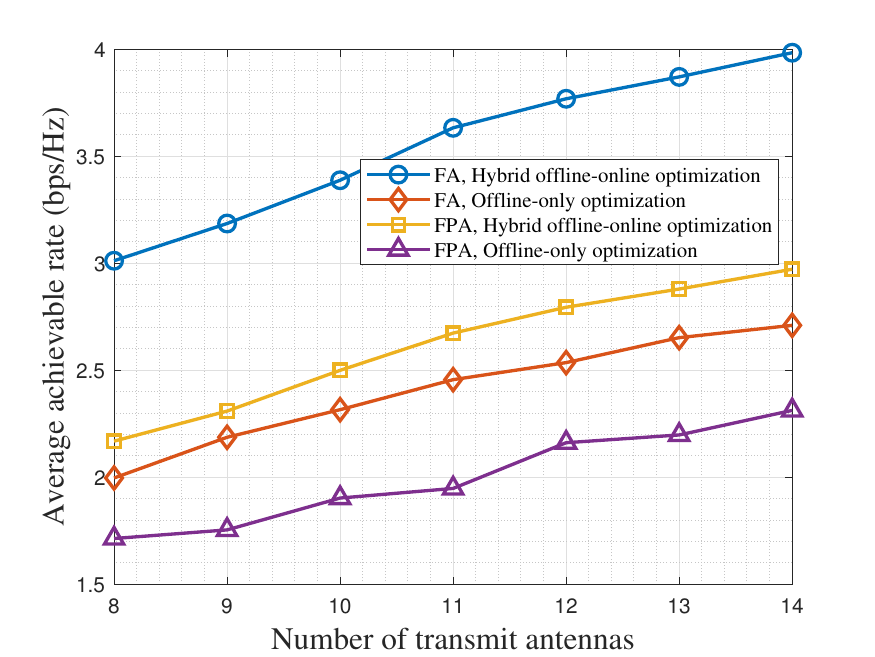}}
\caption{The average achievable rate versus the number of transmit antennas for FA/FPA-enabled UAV systems under offline-only and hybrid offline-online optimization.}
\label{fig:5}
\end{figure}

To demonstrate the proposed framework, we consider the UAV-BS system introduced in Section I, where an FA-enabled UAV directly serves multiple moving ground vehicles. The vehicles follow structured road segments, such as scheduled bus routes, allowing their long-term mobility to be statistically predicted. Before the mission, the UAV has access only to predicted vehicle trajectories obtained from historical data, whereas the actual vehicle locations and instantaneous CSI are available during operation.

As illustrated in Fig.~\ref{fig:4}, the UAV flies from a prescribed initial location to a given final location. The dashed and solid curves denote the predicted and actual vehicle trajectories, respectively. Although mobility uncertainty causes deviations between them, the deviations remain relatively limited because the vehicles follow structured road paths. This setting captures the mismatch between statistically predictable long-term mobility and real-time vehicle locations and channel conditions.

In the offline stage, the predicted vehicle trajectories and probabilistic LoS channel model are used to design a reference UAV trajectory, candidate FA position sets, and a reference beamforming policy. These offline outputs provide the baseline and reduced search space for online optimization. During operation, the UAV adapts its speed according to real-time vehicle locations and instantaneous CSI, while the FA-enabled UAV additionally updates its FA positions and beamforming vectors. The FPA-enabled UAV follows the same framework but updates only its speed and beamforming because antenna positions are fixed.

To evaluate the effectiveness of the proposed hybrid offline-online optimization framework in the FA-enabled UAV-assisted vehicular communication scenario shown in Fig.~\ref{fig:4}, we compare the proposed scheme with three representative schemes, and the four schemes are described as follows:
\begin{itemize}
\item \textbf{FA, hybrid offline-online optimization:} The UAV uses the offline reference trajectory and updates its flight speed, FA positions, and beamforming vectors online.
\item \textbf{FA, offline-only optimization:} The UAV follows the offline-optimized trajectory and uses the offline-designed FA positions and beamforming strategy without real-time adaptation.
\item \textbf{FPA, hybrid offline-online optimization:} The UAV uses an FPA array and updates its flight speed and beamforming vectors online.
\item \textbf{FPA, offline-only optimization:} The UAV uses an FPA array and only relies on the offline-optimized trajectory and beamforming strategy.
\end{itemize}

Fig.~\ref{fig:5} shows that the average achievable rate increases with the number of transmit antenna elements for all schemes due to the additional spatial DoFs and beamforming gains. The FA-enabled schemes consistently outperform their FPA-enabled counterparts because FA reconfiguration provides an additional position-domain DoF: different FA positions induce different effective channel responses and correlations, enabling favorable local channels to be selected and further exploited through beamforming. Thus, the FA gain results from the joint effect of local channel selection and beamforming adaptation.

The hybrid offline-online schemes also outperform their offline-only counterparts because real-time speed and beamforming adaptation can compensate for mismatches between offline predictions and the actual communication environment. The FA-enabled hybrid offline-online scheme achieves the highest average achievable rate, demonstrating the complementary gains of UAV trajectory planning, FA position reconfiguration, and real-time beamforming adaptation.

\section{Future Research Directions}
Beyond conventional A2G data transmission, FA-enabled UAV communication systems can be extended to ISAC, secure communication, MEC, and aerial relaying. In these applications, the coupled optimization complexity and information uncertainty identified in Section III become even more pronounced, motivating application-specific offline-online designs.

\subsection{FA-Enabled UAV-ISAC Systems}
UAV-ISAC systems can simultaneously support communication and sensing services for applications such as target detection, low-altitude mapping, and traffic sensing. With FAs, UAV trajectory planning changes the observation geometry, while FA position reconfiguration adjusts the local spatial channel, providing additional DoFs for jointly improving communication reliability and sensing accuracy. The main challenge is the strong coupling among UAV trajectory, FA positions, and communication/sensing beamforming under conflicting objectives such as spectral efficiency, detection probability, and localization accuracy.

To address this challenge, an offline-online design can exploit environment maps, target priors, and statistical CSI to preplan reference trajectories, candidate FA position sets, and beamforming policies, thereby reducing the real-time optimization space. During operation, instantaneous CSI and real-time echo feedback can be used to locally update FA positions and communication/sensing beamforming. Developing low-complexity algorithms that efficiently balance communication and sensing performance under imperfect prior information remains an important research direction.

\subsection{FA-Enabled UAV Secure Communication Systems}
The LoS-dominant nature of A2G propagation exposes UAV transmissions to potential eavesdroppers over a wide geographical area. FAs can exploit position-dependent channel variations to strengthen legitimate links while weakening eavesdropping links, thereby providing an additional DoF for physical-layer security (PLS). Future designs may jointly optimize UAV trajectory, FA positions, beamforming, artificial noise (AN), and transmit power allocation to improve secrecy performance, which further increases the coupling and computational complexity discussed in Section III.

Meanwhile, eavesdropper locations and CSI are often only partially known, creating substantial information uncertainty. An offline stage can exploit environment maps, historical link feedback, and statistical CSI to identify high-risk regions and construct reference trajectories, candidate FA configurations, and conservative beamforming policies. The online stage can then use real-time legitimate CSI and suspected eavesdropping activity to adapt FA positions, beamforming, and AN power within the reduced decision space. Low-complexity robust designs under imperfect CSI, particularly with multiple eavesdroppers of unknown locations, remain a key challenge.

\subsection{FA-Enabled UAV-MEC Systems}
UAV-mounted MEC systems can provide flexible task offloading, distributed data processing, and intelligent inference services for temporary hotspots, disaster recovery, and other low-altitude applications. Their performance, however, is highly sensitive to unstable A2G links, which may increase transmission latency and energy consumption. FA position reconfiguration can improve uplink offloading and downlink result transmission by selecting favorable channel responses, thereby reducing communication energy and latency and leaving more onboard resources for computation.

The resulting design requires the joint optimization of task offloading, computation resource allocation, UAV trajectory planning, and FA positions under stringent latency and energy constraints, further increasing the variable coupling. To reduce real-time complexity, prior information such as user mobility statistics, service-demand distributions, and statistical channel models can be used offline to determine reference trajectories, candidate FA position sets, and resource allocation policies. Real-time CSI and service-demand information can then support local updates of communication and computing decisions during operation. An important challenge is to balance the communication gain of frequent FA reconfiguration against its signaling and repositioning overhead while adapting to dynamic service demands.

\subsection{FA-Enabled UAV Relay Systems}
UAV relays can provide flexible aerial forwarding for long-distance or blocked links and areas with insufficient infrastructure coverage. Combining UAV trajectory planning with FA position reconfiguration enables relay UAVs to improve the large-scale two-hop propagation geometry while selecting favorable local receiving and forwarding channels. Future research may jointly optimize UAV trajectory, FA positions, relay mode, transmit power control, and beamforming. In single-UAV relaying, the source-to-relay and relay-to-destination links must be balanced, whereas multi-UAV relaying additionally involves node cooperation, interference management, and distributed optimization.

These additional variables and multi-hop channel dependencies further aggravate the computational complexity and information uncertainty. An offline stage can exploit environment maps and statistical CSI to preplan relay trajectories, candidate FA configurations, and transmission policies, while an online stage can use instantaneous CSI to locally update FA positions, power allocation, beamforming, and limited mobility decisions. Such a decomposition can reduce the overhead of global real-time optimization while adapting to time-varying two-hop channels and network topologies. Efficient CSI acquisition and lightweight distributed online optimization remain important issues for large-scale multi-UAV relay networks.

\section{Conclusion}
In this article, we have presented a system-level perspective on FA-enabled UAV communication systems based on dual mobility. UAV trajectory planning provides macro-mobility for large-scale A2G geometry adaptation, whereas FA position reconfiguration provides micro-mobility for fine-grained local channel adaptation. 
Their integration introduces two key challenges: the computational complexity arising from the coupled optimization of UAV trajectory, FA positions, and beamforming, and the information uncertainty and timescale mismatch between prior and real-time information.
To address these challenges, we have presented a hybrid offline-online optimization framework that exploits prior information for long-term planning and real-time information for low-complexity local adaptation. 
The UAV-assisted vehicular communication case study demonstrates the complementary benefits of FA reconfiguration and online adaptation. Future research can further extend this dual-mobility and offline-online design paradigm to ISAC, secure communication, MEC, and aerial relaying.

\bibliographystyle{IEEEtran}
\bibliography{myref}

@ARTICLE{11503149,
  author={Wu, Jun and Yang, Yaoqi and Yuan, Weijie and Liu, Wenchao and Wang, Jiacheng and Mao, Tianqi and Zhou, Lin and Cui, Yuanhao and Liu, Fan and Sun, Geng and others},
  journal={China Commun.}, 
  title={Low-altitude wireless networks: A comprehensive survey}, 
  year={Mar. 2026},
  volume={23},
  number={3},
  pages={99-141},
  doi={10.23919/JCC.fa.2025-0429.202603}}

@ARTICLE{10286328,
  author={Zhu, Lipeng and Ma, Wenyan and Zhang, Rui},
  journal={IEEE Commun. Mag.}, 
  title={Movable Antennas for Wireless Communication: Opportunities and Challenges}, 
  year={Jun. 2024},
  volume={62},
  number={6},
  pages={114-120},
  doi={10.1109/MCOM.001.2300212}}

@ARTICLE{10909643,
  author={Lu, Wen-Jun and He, Chun-Xing and Zhu, Yongxu and Tong, Kin-Fai and Wong, Kai-Kit and Shin, Hyundong and Cui, Tie Jun},
  journal={IEEE Commun. Mag.}, 
  title={Fluid Antennas: Reshaping Intrinsic Properties for Flexible Radiation Characteristics in Intelligent Wireless Networks}, 
  year={May 2025},
  volume={63},
  number={5},
  pages={40-45},
  doi={10.1109/MCOM.002.2400490}}

@ARTICLE{11443232,
  author={Liu, Wenchao and Zhang, Xuhui and Wang, Chunjie and Ren, Jinke and Yuan, Weijie and You, Changsheng},
  journal={IEEE Internet Things Mag.}, 
  title={Fluid Antenna Systems Meet Low-Altitude Wireless Networks: Fundamentals, Opportunities, and Future Directions}, 
  year={May 2026},
  volume={9},
  number={3},
  pages={55-62},
  doi={10.1109/MIOT.2026.3656453}}

@ARTICLE{11214460,
  author={Li, Kaixuan and Yu, Kan and Ma, Dingyou and Zhao, Yujia and Liu, Xiaowu and Zhang, Qixun and Feng, Zhiyong},
  journal={IEEE Trans. Mob. Comput.}, 
  title={Can Movable Antenna-Enabled Micro-Mobility Replace {UAV}-Enabled Macro-Mobility? A Physical Layer Security Perspective}, 
  year={Mar. 2026},
  volume={25},
  number={3},
  pages={4317-4330},
  doi={10.1109/TMC.2025.3624340}}

@ARTICLE{10354003,
  author={Zhu, Lipeng and Ma, Wenyan and Ning, Boyu and Zhang, Rui},
  journal={IEEE Trans. Wirel. Commun.}, 
  title={Movable-Antenna Enhanced Multiuser Communication via Antenna Position Optimization}, 
  year={Jul. 2024},
  volume={23},
  number={7},
  pages={7214-7229},
  doi={10.1109/TWC.2023.3338626}}

@ARTICLE{11302793,
  author={Wu, Tuo and Zhi, Kangda and Yao, Junteng and Lai, Xiazhi and Zheng, Jianchao and Niu, Hong and Elkashlan, Maged and Wong, Kai-Kit and Chae, Chan-Byoung and Ding, Zhiguo and others},
  journal={IEEE Wirel. Commun.}, 
  title={Fluid Antenna Systems Enabling {6G}: Principles, Applications, and Research Directions}, 
  year={2026},
  volume={33},
  number={4},
  pages={100-108},
  doi={10.1109/MWC.2025.3629597}}

@ARTICLE{9034493,
  author={You, Changsheng and Zhang, Rui},
  journal={IEEE Trans. Wirel. Commun.}, 
  title={Hybrid Offline-Online Design for {UAV}-Enabled Data Harvesting in Probabilistic {LoS} Channels}, 
  year={Jun. 2020},
  volume={19},
  number={6},
  pages={3753-3768},
  doi={10.1109/TWC.2020.2978073}}

@ARTICLE{8698468,
  author={You, Changsheng and Zhang, Rui},
  journal={IEEE Trans. Wirel. Commun.}, 
  title={{3D} Trajectory Optimization in {Rician} Fading for {UAV}-Enabled Data Harvesting}, 
  year={Jun. 2019},
  volume={18},
  number={6},
  pages={3192-3207},
  doi={10.1109/TWC.2019.2911939}}

@ARTICLE{10654366,
  author={Liu, Wenchao and Zhang, Xuhui and Xing, Huijun and Ren, Jinke and Shen, Yanyan and Cui, Shuguang},
  journal={IEEE Wirel. Commun. Lett.}, 
  title={{UAV}-Enabled Wireless Networks With Movable-Antenna Array: Flexible Beamforming and Trajectory Design}, 
  year={Mar. 2025},
  volume={14},
  number={3},
  pages={566-570},
  doi={10.1109/LWC.2024.3451246}}

@ARTICLE{11240557,
  author={Zhang, Xuhui and Liu, Wenchao and Ren, Jinke and Wang, Chunjie and Xing, Huijun and Shen, Yanyan and Cui, Shuguang},
  journal={IEEE Trans. Netw. Sci. Eng.}, 
  title={Movable-Antenna Empowered {AAV}-Enabled Data Collection Over Low-Altitude Wireless Networks}, 
  year={2026},
  volume={13},
  number={},
  pages={4506-4523},
  doi={10.1109/TNSE.2025.3631669}}

@ARTICLE{11513706,
  author={Zhu, Xusheng and Wong, Kai-Kit and Hong, Hanjiang and Xiao, Han and Xu, Hao and Wu, Tuo and Chae, Chan-Byoung},
  journal={IEEE Trans. Wirel. Commun.}, 
  title={Fluid Antenna System-Enabled {UAV} Communications in the Finite Blocklength Regime}, 
  year={2026},
  volume={25},
  number={},
  pages={16714-16729},
  doi={10.1109/TWC.2026.3688660}}

@ARTICLE{10065443,
  author={Tian, Kaiyuan and Chi, Yani and Duo, Bin and Yuan, Xiaojun},
  journal={IEEE Commun. Lett.}, 
  title={Hybrid Offline-Online Design for Reconfigurable Intelligent Surface Aided {UAV} Communication}, 
  year={May 2023},
  volume={27},
  number={5},
  pages={1372-1376},
  doi={10.1109/LCOMM.2023.3255241}}

@ARTICLE{11119811,
  author={Yao, Junteng and Wu, Tuo and Zhou, Liaoshi and Jin, Ming and Huang, Chongwen and Yuen, Chau},
  journal={IEEE Trans. Wirel. Commun.}, 
  title={{FAS} Versus {ARIS}: Which Is More Important for {FAS-ARIS} Communication Systems?}, 
  year={2026},
  volume={25},
  number={},
  pages={2075-2091},
  doi={10.1109/TWC.2025.3594617}}

@ARTICLE{11271535,
  author={Wu, Mengru and Lai, Shizhen and Lu, Weidang and Zhao, Bo and Guo, Lei and Nallanathan, Arumugam},
  journal={IEEE Trans. Netw. Sci. Eng.}, 
  title={Rethinking Fluid Antennas for {ISAC}: A Systematic Review and a Novel Air–Ground Collaborative Framework}, 
  year={2026},
  volume={13},
  number={},
  pages={3682-3700},
  doi={10.1109/TNSE.2025.3638847}}

\end{document}